# Engineering of a Layered Ferromagnet *via* Graphitization: An Overlooked Polymorph of GdAlSi

Dmitry V. Averyanov,[a] Ivan S. Sokolov,[a] Alexander N. Taldenkov,[a] Oleg E. Parfenov,[a] Konstantin V. Larionov,[b] Pavel B. Sorokin,[b] Oleg A. Kondratev,[a] Andrey M. Tokmachev,[a] Vyacheslav G. Storchak [a,*]

[a] National Research Center "Kurchatov Institute", Kurchatov Sq. 1, 123182 Moscow, Russia

[b] Laboratory of Digital Materials Science, National University of Science and Technology MISIS, Leninskiy prospect 4, 119049 Moscow, Russia

ABSTRACT – Layered magnets are stand-out materials because of their range of functional properties that can be controlled by external stimuli. Regretfully, the class of such compounds is rather narrow, prompting the search for new members. Graphitization – stabilization of layered graphitic structures in the 2D limit – is being discussed for cubic materials. We suggest the phenomenon to extend beyond cubic structures; it can be employed as a viable route to a variety of layered materials. Here, the idea of graphitization is put into practice to produce a new layered magnet, GdAlSi. The honeycomb material, based on graphene-like layers AlSi, is studied both experimentally and theoretically. Epitaxial films of GdAlSi are synthesized on silicon; the critical thickness for the stability of the layered polymorph is around 20 monolayers. Notably, the layered polymorph of GdAlSi demonstrates ferromagnetism stemming from the open 4$f$-shells of Gd, in contrast to the non-layered, tetragonal polymorph. The ferromagnetism is further supported by electron transport measurements revealing negative magnetoresistance and the anomalous Hall effect. The results show that graphitization can be a powerful tool in the design of functional layered materials.

**INTRODUCTION**

Layered magnets constitute an important class of compounds renowned for their unconventional physical properties.[1,2] The current interest to such materials is spurred by the discovery of 2D magnetism in a few monolayers (ML) of layered materials.[3,4] 2D magnets hold high promise for new quantum phases and applications in ultra-compact spintronics.[5-7] Their functionality benefits from high amenability of 2D magnetism to external stimuli such as magnetic[3] and electric[8] fields or pressure.[9] This property extends to multilayer materials; in particular, pressure may readily control the interlayer magnetic coupling.[10] In layered magnets, magneto-optical phenomena – linear magnetic dichroism,[11] helical light emission,[12] magneto-optical Kerr effect[13] – can be remarkably enhanced.[14] Rich phase diagrams are the norm for this class of materials often exhibiting a variety of magnetic quasiparticles[15-18] and complex magnetic orders[19] that can be strongly thickness-dependent.[20,21] Topological states in layered magnets attract much attention as well.[22,23] Taking into account all these advantageous properties, significant efforts are invested into attempts to extend the pool of layered magnets, in particular, to materials based on 4*f*-elements with promising topological structures.[24-27] However, the material landscape of layered magnets is still scarcely populated, making a case for design of new materials and synthetic methods. An additional challenge is that technological considerations require the materials to be in the form of nanoscale films, with good conductivity, epitaxially integrated with mature semiconductor platforms.

Ever since the rise of graphene, the honeycomb lattice has been inspiring research in materials science.[28-31] Regretfully, the number of layered materials with the honeycomb structure is relatively small, a pool of magnetic systems being even smaller. To avoid searches in the dark, design and synthesis of new honeycomb materials require some guiding principles. A possible solution is to explore the rich toolbox of polymorph engineering[32] – formation of a different crystal structure without changes to the material composition. One option to make honeycomb materials is by employing the concept of spontaneous graphitization:[33] it has been predicted that a number of cubic lattices, known in the bulk, prefer a layered graphitic structure below some critical thickness, on the order of a dozen of ML.[34] As an example, graphitic-like hexagonal phases of alkali halides, proposed in Refs. [33,34], have been produced by confinement[35] or deposition on a specially chosen substrate.[36] The concept of pseudomorphism suggests that unusual phases can be stabilized by a substrate in the process of epitaxy.[37] Graphitization is driven by the surface energy; its contribution to the total energy gains weight as the film thickness decreases. The surface energy of a cubic material is generally higher than that of its layered counterpart.[33] Graphitization may be helpful in synthesis of magnetic materials with the honeycomb structure. In particular, hexagonal phases of chromium nitride[38] and manganese nitride,[39] promising for spintronic applications, have been predicted to be stable in the 2D limit but the conclusions are yet to find experimental support. Looking from a more general perspective, the question is whether the phenomenon of graphitization is general, whether it extends beyond cubic structures.

Design of a magnet normally requires the presence of elements with open shells. The majority of the layered magnets are based on *d*-elements. However, lanthanide materials can often provide a viable alternative – they demonstrate advantageous functional properties such as high carrier mobility[40,41] or colossal negative magnetoresistance,[16] can be applied as drug delivery systems,[42] photoluminescent materials or catalysts.[43] Brought to the 2D limit, rare-earth materials exhibit great potential in optoelectronics, design of transistors and magnetic devices.[44] The 2D magnet $EuSi_2$ is a prominent member of the class[16,20] with a honeycomb structure provided by layers of silicene, the Si counterpart of graphene.[45] The polymorphism of $EuSi_2$ may be understood in terms of graphitization: $EuSi_2$ has usually a tetragonal lattice with a 3D net of Si atoms[46] but its trigonal silicene-based polymorph (unknown in the bulk) can be stabilized in the limit of ultrathin films.[47] As soon as the film thickness exceeds about a dozen ML the trigonal polymorph transforms into the tetragonal $EuSi_2$. Our goal is to use this pattern to design a new layered magnet. The magnetism of Eu-based compounds stems from the $4f^7$ configuration of the cation $Eu^{2+}$. The same configuration is provided by $Gd^{3+}$ cations. The problem is that Eu and Gd donate different numbers of electrons into the anionic system. However, we can exploit the idea of isoelectronicity, useful in materials science:[48,49] replacement of half of Si atoms with Al to compensate the extra electrons from Gd suggests a material of the GdAlSi stoichiometry (see Figures 1a,b) isoelectronic to $EuSi_2$ (in both compounds the valence electron count is equal to 10). The graphene-like structure of the AlSi anionic layer is well known in CaAlSi[50] and SrAlSi,[51] studied extensively for their superconducting properties. It should also be noted that rare-earth materials with different graphene-like anionic layers are established as candidates for realizing magnetically tuned topological properties.[52] In the bulk, GdAlSi crystallizes in a tetragonal structure,[53] just as $EuSi_2$ does. However, as in the case of $EuSi_2$, we can rely upon the power of graphitization to produce an overlooked polymorph of GdAlSi with a honeycomb structure.

Here, we present experimental and theoretical studies of GdAlSi films on silicon. We demonstrate graphitization of GdAlSi in the 2D limit resulting in a new, layered, magnetic material, determine the critical thickness, and devise synthetic routes to different polymorphs. The syntheses are accompanied by an extensive study of the atomic and electronic structures, magnetic and electron transport properties. In particular, we put emphasis on the ferromagnetic properties of the new polymorph of GdAlSi setting it apart from the mundane tetragonal polymorph.

**RESULTS AND DISCUSSION**

**Synthesis of Layered GdAlSi**

Synthesis of new polymorphs is usually accomplished by application of external pressure or heating to high temperature. In contrast, our aim is to engineer a layered polymorph of GdAlSi by manipulating the interfacial contribution to the total energy. First, the role of the interfacial energy should be significant – therefore, we concentrate our efforts on ultrathin films comprising no more than a few dozen ML. It also affects our choice of the synthesis technique: we employ molecular beam epitaxy (MBE) – a versatile technique providing control over the film growth on the level of a single ML. Second, the material is ternary meaning that supplies of 3 elements – Gd, Al, and Si – should be balanced to provide the necessary stoichiometry. Otherwise, side products are expected because the phase diagram Gd-Al-Si is rich: it includes such compounds as $GdAl_2Si_2$, $GdAl_n$ ($n$ = 1–3), $Gd_3Al_2$, various Gd silicides. Therefore, keeping the balance for all the 3 elements would be rather challenging. To avoid this problem, we employ silicon as a substrate to synthesize GdAlSi. In this case, the substrate serves also as a reactant; the right amount of Si, controlled by the fluxes of Gd and Al, is supplied to the surface *via* the efficient vacancy mechanism.[20,47,51] Third, the substrate should stabilize the intended polymorph structure, *i.e.* have the same symmetry and matching lattice parameters. Therefore, we employ the (111) face of silicon: its surface represents a covalently bound silicene of the honeycomb structure compatible with the graphene-like AlSi layer.

To synthesize the layered (trigonal) polymorph of GdAlSi we remove the natural oxide from the Si(111) surface to produce the characteristic $7 \times 7$ reconstruction and then deposit stoichiometric amounts of Gd and Al at the heated substrate (see Methods). It should be noted that the synthesis does not require any template layers; this is in contrast to MBE synthesis of SrAlSi on Si(111)[51] involving a self-sacrificial template of $SrSi_2$ for epitaxy. The procedure results in the layer-by-layer growth of the monocrystalline trigonal polymorph. However, the situation changes dramatically as soon as the film thickness exceeds some critical value, around 20 ML – another, tetragonal, polymorph of GdAlSi emerges making the film polycrystalline. This transformation is consistent with the nature of the graphitization phenomenon: as soon as the surface energy does not contribute a significant part of the total energy, the most stable polymorph is that corresponding to the ground state of the bulk. To carry out further studies, we synthesized a set of films with a thickness about 10 ML, reasonably far from the critical value. Also, for comparative purposes, we produced a matching sample of the tetragonal polymorph stable in the bulk. Its synthesis was carried out on Si(001) which surface stabilizes the tetragonal lattice, as in the case of tetragonal $EuSi_2$.[46] The conditions of the synthesis were exactly the same as in the case of trigonal GdAlSi, the only difference being the face of the substrate. To protect the samples from oxidation by air, they were capped by a layer of amorphous $SiO_x$, a non-magnetic insulator that does not affect the results of our *ex situ* studies of the atomic structure, magnetic and transport properties.

**Atomic Structure of GdAlSi**

As the trigonal polymorph of GdAlSi has not been reported before, its structural parameters are unknown. To analyze the atomic structure, we use a combination of experimental and computational techniques. First, the structure of the material is monitored in the MBE growth chamber employing reflection high-energy electron diffraction (RHEED), a technique sensitive to the state of the surface. A characteristic RHEED image of the GdAlSi film, presented in Figure 1d, agrees with the trigonal structure of GdAlSi shown schematically in Figures 1a,b. The lateral lattice parameter determined from the peak positions in the RHEED image is $a = 4.03(3)$ Å. The difference with the corresponding lattice parameter of the Si(111) surface ($a = 3.84$ Å) is about 5% – the lattice mismatch is sufficiently small for successful epitaxy. Also, the lattice parameter *a* differs significantly from that in $GdSi_2$ ($a = 3.894(2)$ Å[54]), a silicide phase similar to GdAlSi. A possible reason is that the flat anionic layer in $GdSi_2$ is stabilized by a significant amount of charge-balancing vacancies (shrinking the lattice) whereas the same balance in GdAlSi is achieved by replacing a half of Si with Al.

A further structural study of the material was carried out *ex situ* employing X-ray diffraction (XRD). Figure 1e shows a θ-2θ XRD scan of an 11 ML sample of trigonal GdAlSi on Si(111). It manifests the epitaxial character of the film and the absence of any side phases for this particular thickness below the critical value. Moreover, the well-developed thickness fringes of the diffraction peaks point at a high quality of the film and its interfaces. An estimate of the film thickness based on XRD oscillations, 11.3 ML, agrees well with the nominal value, 11 ML, corresponding to the time of Gd and Al deposition. The thickness of the film is further corroborated by X-ray reflectivity analysis. The parameter $c$ of trigonal GdAlSi, calculated from the positions of the (000$n$) XRD reflexes, is 4.148(5) Å. In EuSi$_2$, the anionic silicene layers are strongly buckled,[47] as in free-standing silicene. The question is whether the AlSi layer in GdAlSi is buckled, as in its isoelectronic counterpart EuSi$_2$, or flat, as in CaAlSi and SrAlSi. The experiment does not give a direct answer but the value of the lattice parameter $c$ is a strong indication that the AlSi layer is flat. Indeed, the parameter $c$ in GdAlSi is smaller by 20% than $c$ in EuSi$_2$[47] but is close to $c$ in GdSi$_2$,[54] a compound with flat anionic layers. A similar RHEED and XRD study was carried out for tetragonal GdAlSi. The determined lattice parameters, $a = 4.15(3)$ Å and $c = 14.3874(7)$ Å agree with those of bulk GdAlSi, $a = 4.1255(1)$ Å and $c = 14.4259(5)$ Å.[53] The equivalence of the synthetic conditions, including Gd and Al fluxes, for the two polymorphs provides an argument that the trigonal compound has the same GdAlSi stoichiometry as the tetragonal counterpart. However, we cannot exclude some non-stoichiometry in the anionic layer; after all, other compounds with the AlSi layer are berthollides.[55] An experimental approach to answer this question seems hopeless for the ultrathin films. It has been pointed out in Ref. [56] that even in the bulk the actual distribution of Al and Si in the AlSi layer is very difficult to establish experimentally.

To get further insight and support the experimental observations, we provide the results of a thorough study based on density functional theory (DFT), a working horse of computational materials science. The DFT + U calculations, to account for the local electron correlations on Gd, estimate the lattice parameters of trigonal GdAlSi to be $a = 4.17$ Å and $c = 4.00$ Å. The small discrepancy (~ 3%) with the experimental values can be partially attributed to the film thinness and the role of the Si substrate in the experimental setting, *i.e.* effects absent in the bulk crystal simulation. The calculations suggest the anionic layers to be flat as in graphene, not buckled as in silicene: as shown in Figure 1c, the energy minimum for the buckling parameter corresponds to $\Delta = 0$. This calculation agrees with the above conclusion based on the lattice parameter $c$.

The key characteristic of the system to analyze computationally is the interplay between the polymorphs depending on the film thickness. Experimentally, it results in the appearance of both phases, trigonal and tetragonal, for sufficiently thick films. This point is illustrated by inset of Figure 2 – peaks from both polymorphs appear in the XRD θ-2θ scan of a 28 ML film. Theoretically, there should be an energy competition resulting in the change of the ground state polymorph. We calculated the total energy for both polymorphs as a function of the number of ML (in the case of non-layered GdAlSi, 1 ML corresponds to one $Gd_1Al_1Si_1$ unit along the $c$-axis). The results are plotted in Figure 2 as dependences of the total energy per formula unit on the inverse number of ML. These coordinates linearize the dependences as long as the total energy can be well represented as a sum of two contributions – the bulk energy proportional to the number of ML and the nearly constant surface energy. According to Figure 2, the dependences are indeed linear. Most important, the two linear trends intersect – the trigonal polymorph becomes the most stable phase below 18 ML. This critical value nicely corresponds to the experiment; although, the simulation does not take into account the interaction of the film with the substrate. In any case, the calculations predict that the surface energy drives the graphitization process of GdAlSi, in agreement with the experiment.

**Magnetic and Transport Properties of GdAlSi**

As stated above, design and synthesis of the trigonal polymorph of GdAlSi aim at a new layered magnet. Therefore, studies of the GdAlSi magnetism and its influence on electron transport are important. Figure 3 demonstrates basic magnetic properties of trigonal GdAlSi. Temperature dependence of the magnetic moment (Figure 3a) indicates that the material exhibits ferromagnetic (FM) behavior with the characteristic transition temperature $T_C$ in the region 35-40 K. In 2D materials, the effective transition temperature may depend on weak magnetic fields.[4,20,21] In the 11 ML sample, this dependence, if any, is not particularly strong. Magnetic field dependence (*M-H* curve, Figure 3b) shows a significant hysteresis loop shrinking with temperature. At 40 K, above $T_C$, the magnetic moment becomes negligible. The material demonstrates other characteristics typical for the FM ordering. In particular, it exhibits a significant remnant magnetic moment (Figure 3c) decaying as the temperature increases towards $T_C$ as well as bifurcation of the field-cooled (FC) and zero-field-cooled (ZFC) magnetization (Figure 3d).

The magnetic properties of trigonal GdAlSi are highly anisotropic. Figure 4a compares temperature dependence of the magnetic moments for in-plane and out-of-plane magnetic fields. The characteristic increase of $M_{FM}$, associated with the emergence of an FM order, is detected for in-plane magnetic fields only. Thus, GdAlSi exhibits easy-plane magnetic anisotropy. This conclusion is supported by a computational non-collinear magnetic study with spin-orbit coupling: the calculated magnetic anisotropy energy (inset in Figure 4a) suggests that the spins prefer the in-plane orientation. In fact, easy-plane magnetism is rather typical for layered rare-earth compounds with triangular lattices of $4f^7$ ions.[20,21,57-60] It is important that the magnetic state is sensitive to the atomic structure. Figure 4b compares temperature dependence of molar magnetic susceptibility in the 2 polymorphs of GdAlSi: it is featureless in the case of antiferromagnetic (AFM) tetragonal GdAlSi but demonstrates a significant upturn around $T_C$ in the case of the trigonal polymorph. Graphitization of GdAlSi has resulted in qualitative changes of the magnetic state. The situation is similar to the magnetism of $GdSi_2$ polymorphs demonstrated in Ref. [20].

Based on the $4f^7$ configuration of Gd ions, one may expect the magnetic moments to reach 7 $\mu_B$/Gd. Analysis of the *M-H* curves (Figure 3b) reveals that the saturation FM moments are rather low, about 0.2 $\mu_B$/Gd. This is actually not surprising because various magnets with $4f^7$ ions demonstrate reduced magnetic moments.[20,21] A reasonable explanation is that a phase-separated state including both FM and AFM regions sets in. It is supported by XMCD measurements[61] and observation of exchange bias in $GdSi_2$[57] and $EuSi_2$.[58] It should be noted that the reduced magnetic moments are not a consequence of frustration in the triangular lattice of rare-earth ions – the situation is the same for other lattices as well.[62] Also, the reduced moments do not arise from some odd-even effects related to the A-AFM magnetic order: we synthesized a number of films between 9 and 12 ML – their magnetic moments are practically the same.

We carried out spin-polarized calculations for FM and AFM orders. The optimized lattice parameters are found to be largely insensitive to the magnetic configuration. The AFM configuration is lower in energy but the energy difference is only 14 meV per formula unit. This small difference may explain coexistence of FM and AFM states in the material. Figure 5 presents the band structure as well as total and orbital-resolved density of states (DOS) plots for the FM and AFM configurations. The 4*f*-orbitals are far from the Fermi energy; the states around the Fermi energy are dominated by Gd *d* orbitals. Both systems are predicted to be metals. This shouldn't be a surprise because various related magnetic compounds – silicides[16,20] and germanides[21] of Eu and Gd, GdScSi and GdScGe[63] – are metals. The metallic behavior of such Zintl phases is discussed in Ref. [[64]]. Naturally, the results of the computational study require experimental confirmation.

We studied lateral electron transport in trigonal GdAlSi. Figure 6a presents temperature dependence of the film resistivity. The material is a metal – the value of the resistivity $\rho_{xx}$ is rather low. The feature just below 40 K corresponds to the magnetic transition (*cf.* Figures 3a and 4). The upturn of resistivity below 10 K can be associated with Kondo behavior – it is similar to EuSi$_2$ and GdSi$_2$, both suggested to be Kondo metals.[16] According to Figure 6b, the Hall effect in GdAlSi is non-linear below $T_C$. The presence of hysteresis suggests that the non-linearity stems from the anomalous Hall effect. Further support of the magnetism in GdAlSi comes from magnetoresistance (MR) measurements for different directions of the applied magnetic field (Figure 6c). The MR is highly anisotropic; in particular, its sign depends on the direction of the in-plane magnetic field – parallel or perpendicular to the current. In the latter case, the MR is negative and exhibits a quasi-linear behavior. The hysteresis at low magnetic fields (Figure 6d) is yet another manifestation of the GdAlSi magnetism. The negative MR becomes smaller as temperature increases but it doesn't vanish at $T_C$ which may indicate the presence of FM correlations.

**CONCLUSION**

Design of functional materials relies upon our understanding of the stability of compounds, its dependence on the chemical composition and the atomic structure. These considerations are especially important for approaches based on polymorph engineering. We employed two powerful concepts, graphitization and isoelectronicity, to produce a new layered magnet. Graphitization, stabilization of layered polymorphs in the 2D limit, is a natural phenomenon arising from the surface energy lowering in layered structures. Isoelectronicity helps in identifying related compounds which is particularly important for prediction of the structures and properties of new systems based on their known isoelectronic counterparts. In our research, we employed EuSi$_2$, a material exhibiting stabilization of the honeycomb silicene lattice in ultrathin films, as a blueprint. The idea was that the change of Eu by Gd, forming isoelectronic cations, and the replacement of the anionic silicene layer with AlSi, to reach the overall isoelectronicity, would produce a layered magnet. Accordingly, we were able to synthesize a new, layered polymorph of GdAlSi with graphene-like anionic layers AlSi. As expected, the material is stable only below some critical thickness, estimated to be around 20 monolayers. Most important, the new polymorph of GdAlSi exhibits ferromagnetic properties, unlike its non-layered counterpart, stable in the bulk phase. This conclusion is supported by magnetization and electron transport measurements, also indicating competition of magnetic states. The results suggest that graphitization can serve as a general pathway to engineer materials by design. For instance, one can try to replace Si with Ge and produce layered GdAlGe: isoelectronic layered EuGe$_2$ is well known,[65] the bulk structure of GdAlGe is the same as in the case of GdAlSi.[66] Hopefully, the presented design and synthesis of an overlooked layered magnet will not be an isolated example and graphitization will prove itself a powerful instrument in the toolbox of materials science. The ability to manipulate the stability of the polymorphs through the film thickness and the observed magnetic behavior make GdAlSi a compelling candidate for further exploration in the field of functional materials.

**METHODS**

**Synthesis**

The materials were synthesized in a Riber Compact system for molecular beam epitaxy. The syntheses were carried out in ultra-high vacuum corresponding to a base pressure below 10$^{-10}$ Torr. Two types of substrates were used: Si(111) for the trigonal polymorph of GdAlSi and Si(001) for the tetragonal polymorph. In both cases, the miscut angle did not exceed 0.5° and the lateral size of the substrates was 1 inch. The layer of the natural oxide SiO$_x$ was removed from the substrates by heating at 950 °C; the removal resulted in the formation of the standard surface reconstructions – 7 × 7 for Si(111) and 2 × 1 for Si(001). The substrate temperature was determined by a PhotriX ML-AAPX infrared (0.9 μm wavelength) pyrometer. To synthesize the polymorphs of GdAlSi, stoichiometric amounts of 4N Gd and 5N Al were deposited simultaneously on the substrates. The synthesis conditions were identical for the 2 polymorphs:

the Si substrate was kept at 400 °C; the pressures of Gd and Al, measured by a Bayard-Alpert ionization gauge, were $1 \cdot 10^{-8}$ and $6 \cdot 10^{-9}$ Torr, respectively. The beams of Gd and Al were produced by heating the respective Knudsen cell effusion sources to $T_{Gd}$ = 1210 °C and $T_{Al}$ = 905 °C. The films were capped with a 20 nm layer of the amorphous insulator $SiO_x$, deposited at room temperature, to protect them from oxidation by air.

**Characterization**

The structure of the GdAlSi films was established by diffraction techniques, RHEED and XRD. The syntheses were monitored *in situ* employing a RHEED diffractometer furnished with the kSA 400 analytical system. The θ-2θ XRD scans were produced *ex situ* by a Rigaku SmartLab 9 kW diffractometer. They were recorded for a wavelength of 1.54056 Å (Cu $K_{\alpha 1}$). The same diffractometer was employed to carry out X-ray reflectivity analysis of the film thickness. The magnetic moments of the films were measured by an MPMS XL-7 SQUID magnetometer employing the reciprocating sample option (RSO). GdAlSi samples with a lateral size 5 mm were mounted in plastic straws; the precision of the film orientation with respect to the applied magnetic field was better than 2°. To study electron transport in the films, a LakeShore 9709A system was employed. Four-contact measurements were carried out for square samples 5 mm × 5 mm. The electrical contacts to the samples were made by deposition of an Ag-Sn-Ga alloy. The quality of the contacts was attested by measurements of I-V characteristic curves.

**Computational Techniques**

Spin-polarized ground-state calculations of the GdAlSi polymorphs were carried out by the DFT + U method[67] employing the Perdew-Burke-Ernzerhof (PBE) parameterization[68] of the generalized gradient approximation. The parameters *U* and *J* were set to the values 6.7 and 0.7 eV, respectively, to improve the treatment of 4*f* states in Gd atoms.[69] For calculations, we used the projector augmented wave method with periodic boundary conditions, as implemented in the Vienna Ab initio Simulation Package (VASP).[70-72] The plane-wave energy cut-off was set to 400 eV. Dispersion corrections were taken into account employing the D3 scheme.[73] The Brillouin zone was sampled according to the Monkhorst-Pack schemes with a *k*-point spacing of 0.16 Å$^{-1}$ and 0.08 Å$^{-1}$ for the relaxation and density of states calculations, respectively. To avoid spurious interactions in slab calculations, the vacuum size along the non-periodic direction was set to 20 Å. The magnetic anisotropy energy was calculated in a non-collinear magnetic regime. Post-processing was carried out using the VESTA[74] and VASPKIT[75] packages.

AUTHOR INFORMATION

**Corresponding Author**

* E-mail: vgstorchak9@gmail.com (VGS)

**Author Contributions**

The manuscript was written through contributions of all authors. All authors have given approval to the final version of the manuscript.

**Notes**

The authors declare no conflict of interest.

ACKNOWLEDGMENT

This work was supported by NRC "Kurchatov Institute" and the Russian Science Foundation [grants No. 22-13-00004 (synthesis), No. 23-79-01298 (DFT calculations), No. 19-19-00009 (magnetism studies), and No. 20-79-10028 (electron transport studies)]. The Laboratory of Digital Material Science was created with the support by the Ministry of Science and Higher Education of the Russian Federation in the framework of the Increase Competitiveness Program of NUST "MISIS" (grant No. K6-2022-041). D.V.A. acknowledges support from the President's scholarship (SP 3111.2022.5). The measurements were carried out using equipment of the resource centers of electrophysical and laboratory X-ray techniques at NRC "Kurchatov Institute"; the calculations were carried out using a supercomputer cluster provided by the Materials Modeling and Development Laboratory at NUST "MISIS" and the Joint Supercomputer Center of the Russian Academy of Sciences.

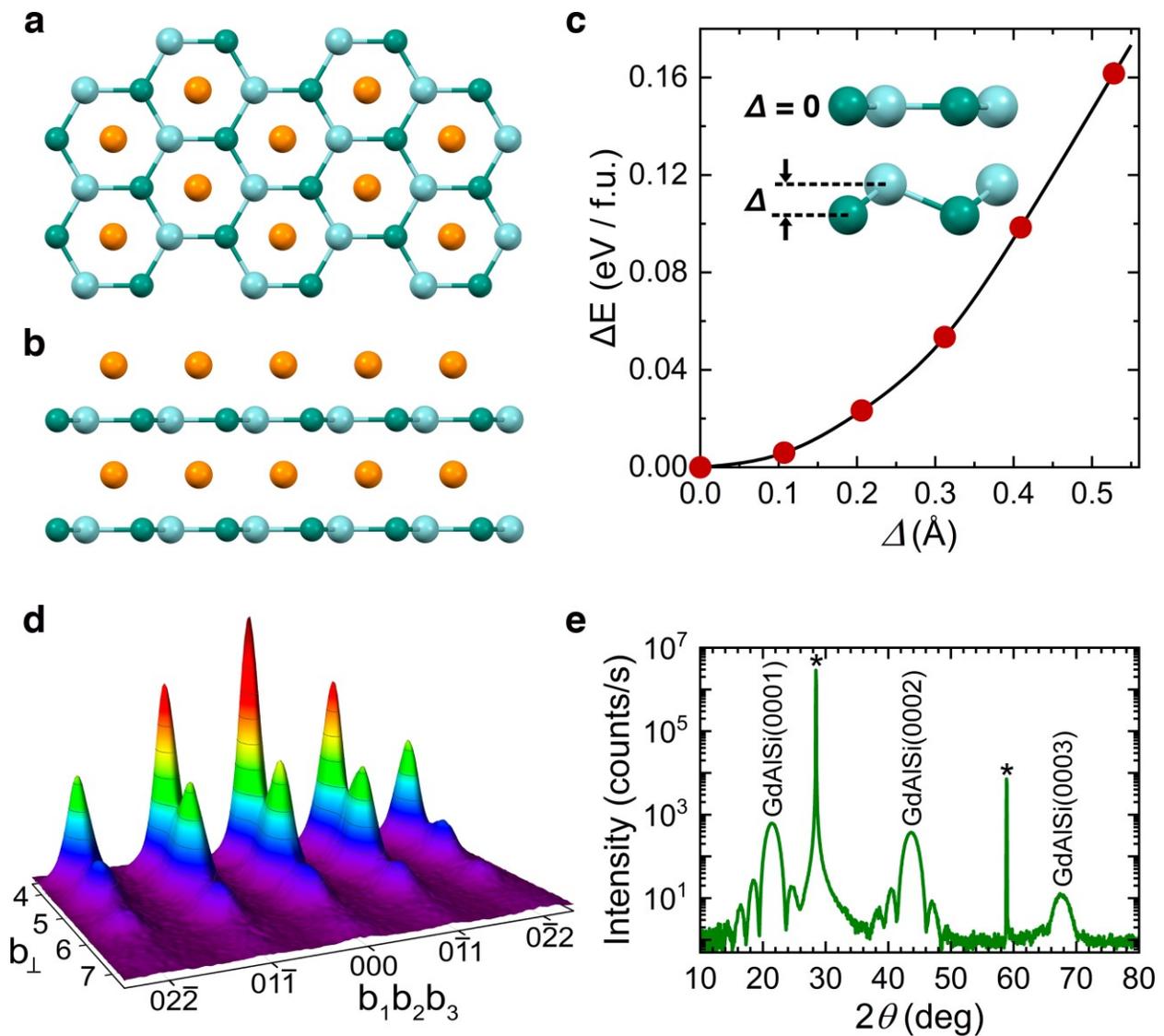

**Figure 1.** a) Top view (honeycomb AlSi layers with Gd atoms in the centers of the hexagons) and b) side view (layered structure) on a ball-and-stick model of the trigonal GdAlSi polymorph (Gd – orange, Al – teal, and Si – green); c) DFT+U calculated dependence of the unit cell energy on the AlSi layer buckling $\Delta$; d) 3D RHEED image of an 11 ML GdAlSi film; the reflexes are marked by 4 Miller-Bravais indices – $b_1b_2b_3$ for the basal plane and $b_\perp$ for the vertical direction; e) θ-2θ XRD scan of an 11 ML GdAlSi film; asterisks denote peaks from the silicon substrate.

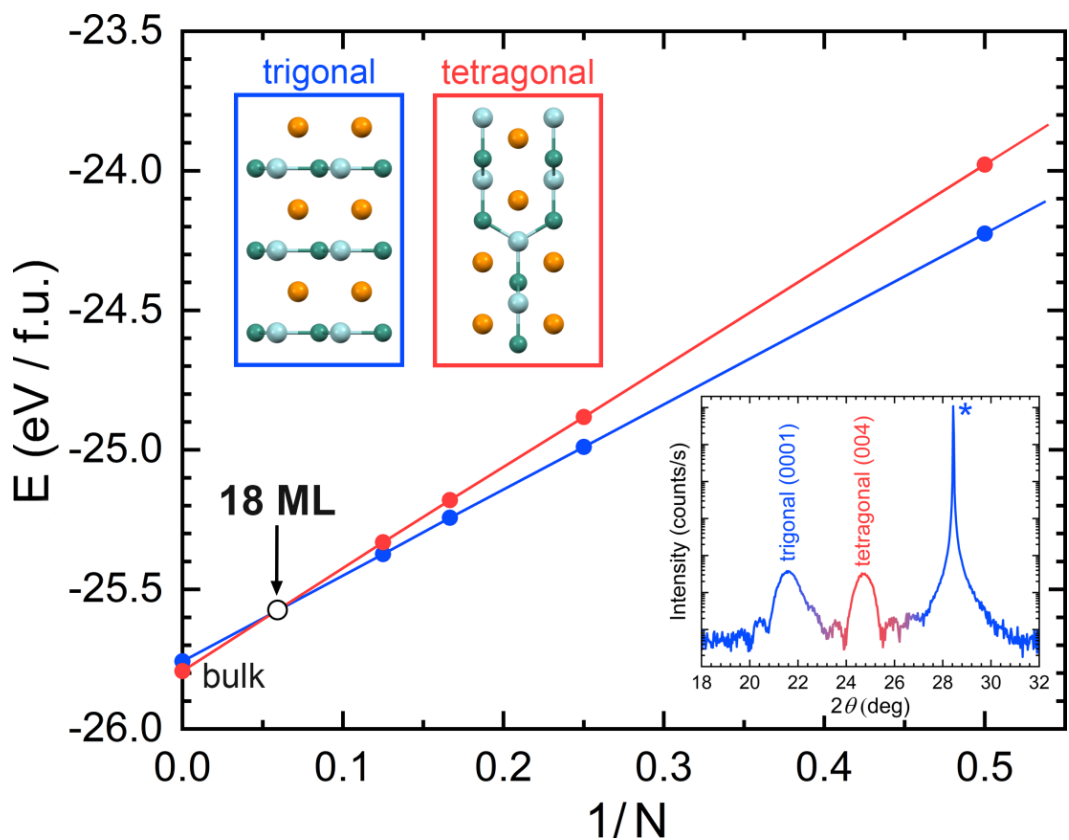

**Figure 2.** Dependence of the total energy (calculated by DFT+U, eV per formula unit) on the inverse number of ML in the trigonal (blue) and tetragonal (red) polymorphs of GdAlSi; top left inset: the structures of the polymorphs; bottom right inset: a fragment of the θ-2θ XRD scan showing peaks from trigonal (blue) and tetragonal (red) polymorphs of GdAlSi in a 28 ML film; asterisk denotes a peak from the Si substrate.

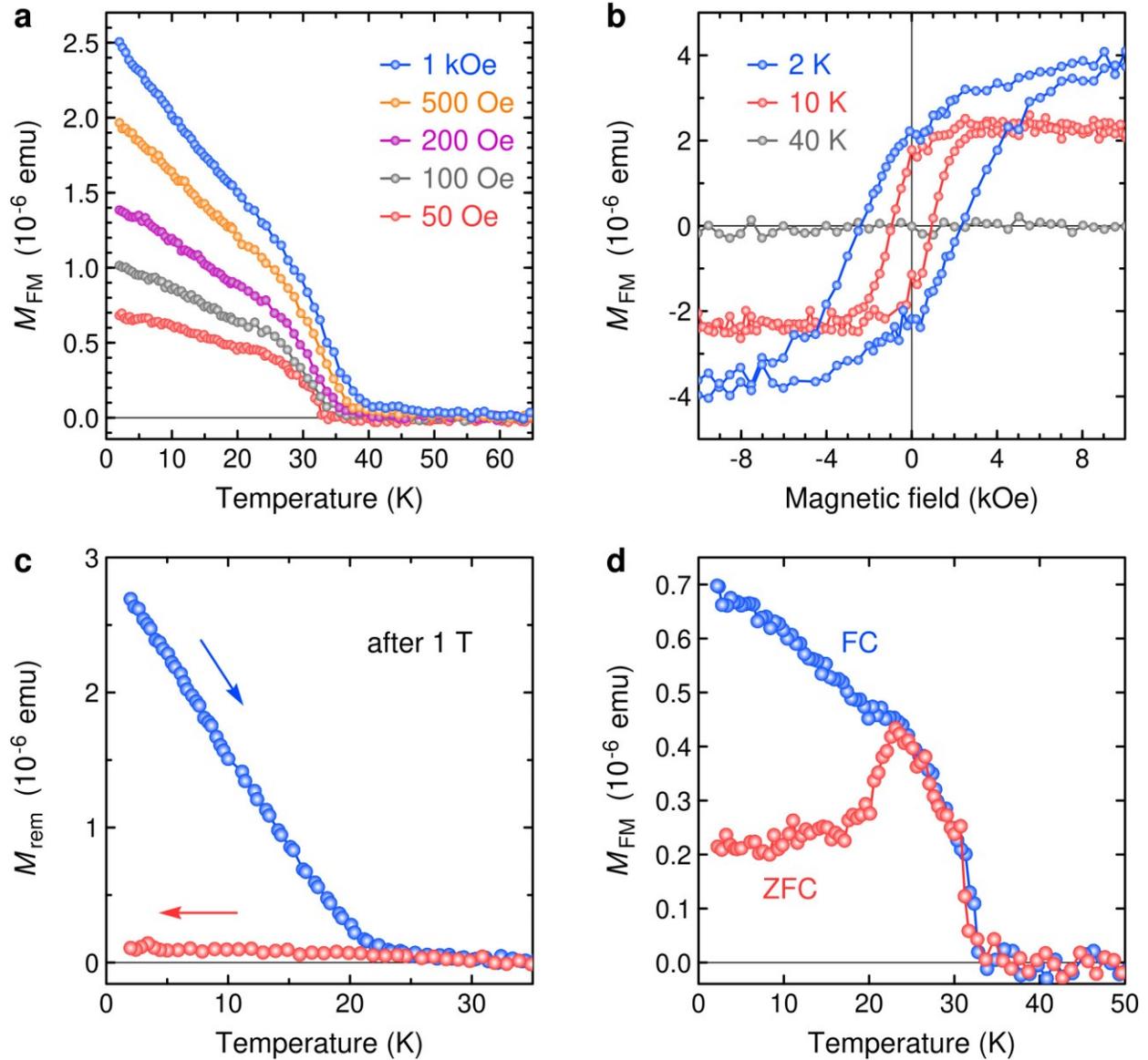

**Figure 3.** Magnetic properties of the trigonal polymorph of GdAlSi (11 ML). a) Temperature dependence of the FM moment in in-plane magnetic fields 50 Oe (red), 100 Oe (grey), 200 Oe (purple), 500 Oe (orange), and 1 kOe (blue); b) in-plane magnetic field dependence of the FM moment at 2 K (blue), 10 K (red), and 40 K (grey); *M-H* hysteresis loops are shown for 2 K and 10 K; c) temperature dependence of the remnant moment after cooling in an in-plane magnetic field 1 T; d) temperature dependence of the FM moments for zero-field cooling (ZFC, red) and field cooling (FC, blue) in an in-plane magnetic field 50 Oe.

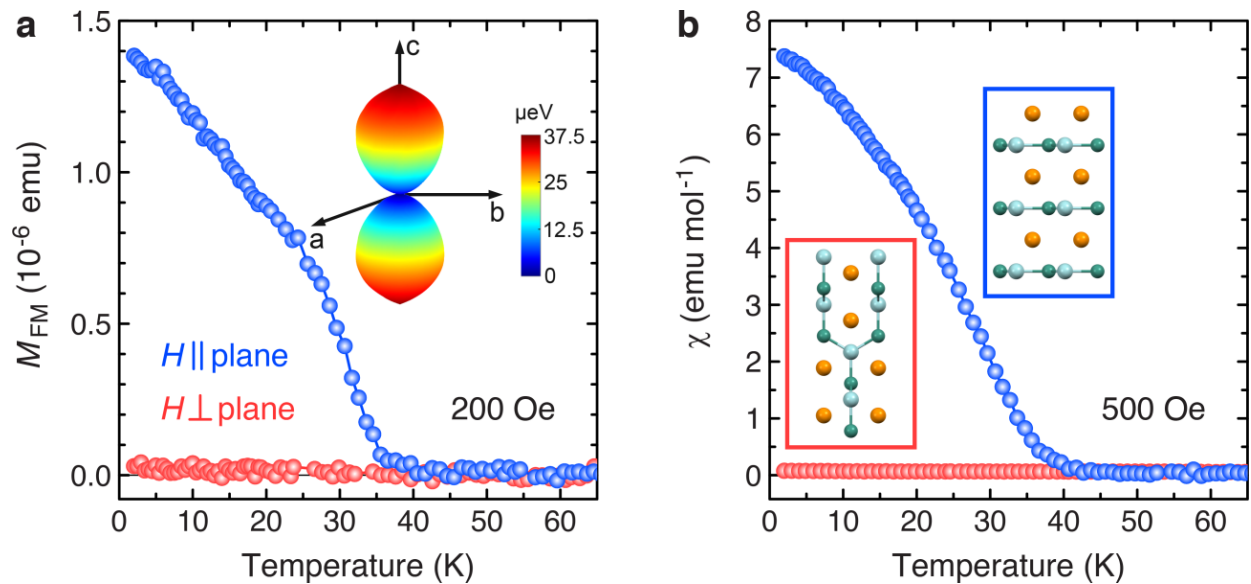

**Figure 4.** a) Temperature dependence of the FM moment in the trigonal polymorph of GdAlSi (11 ML) in in-plane (blue) and out-of-plane (red) magnetic fields 200 Oe; inset: calculated magnetic anisotropy energy (per formula unit) of bulk GdAlSi; b) Temperature dependence of the molar magnetic susceptibility $\chi$ in trigonal (blue) and tetragonal (red) polymorphs of GdAlSi, measured in an in-plane magnetic field 500 Oe.

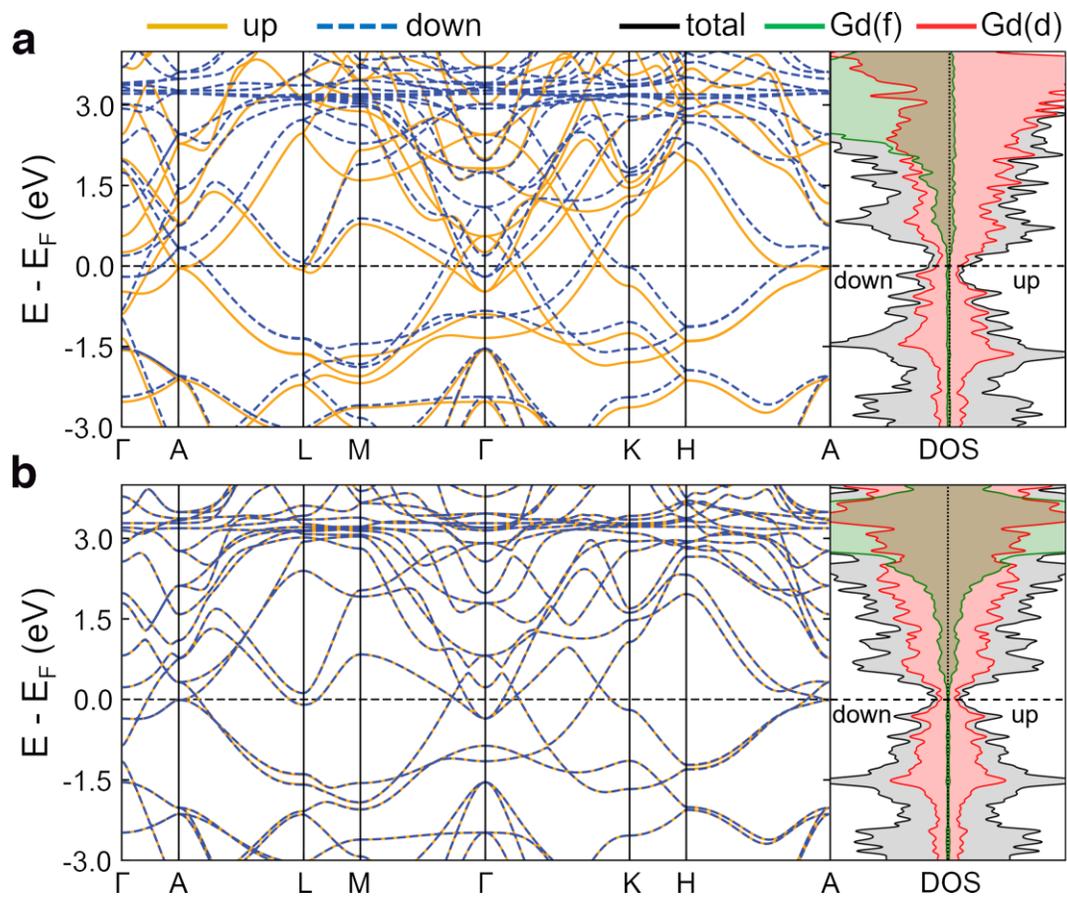

**Figure 5.** Band structure and orbital-resolved DOS of the bulk of the trigonal polymorph of GdAlSi with a) FM and b) AFM ordering. The Fermi energy is shifted to zero.

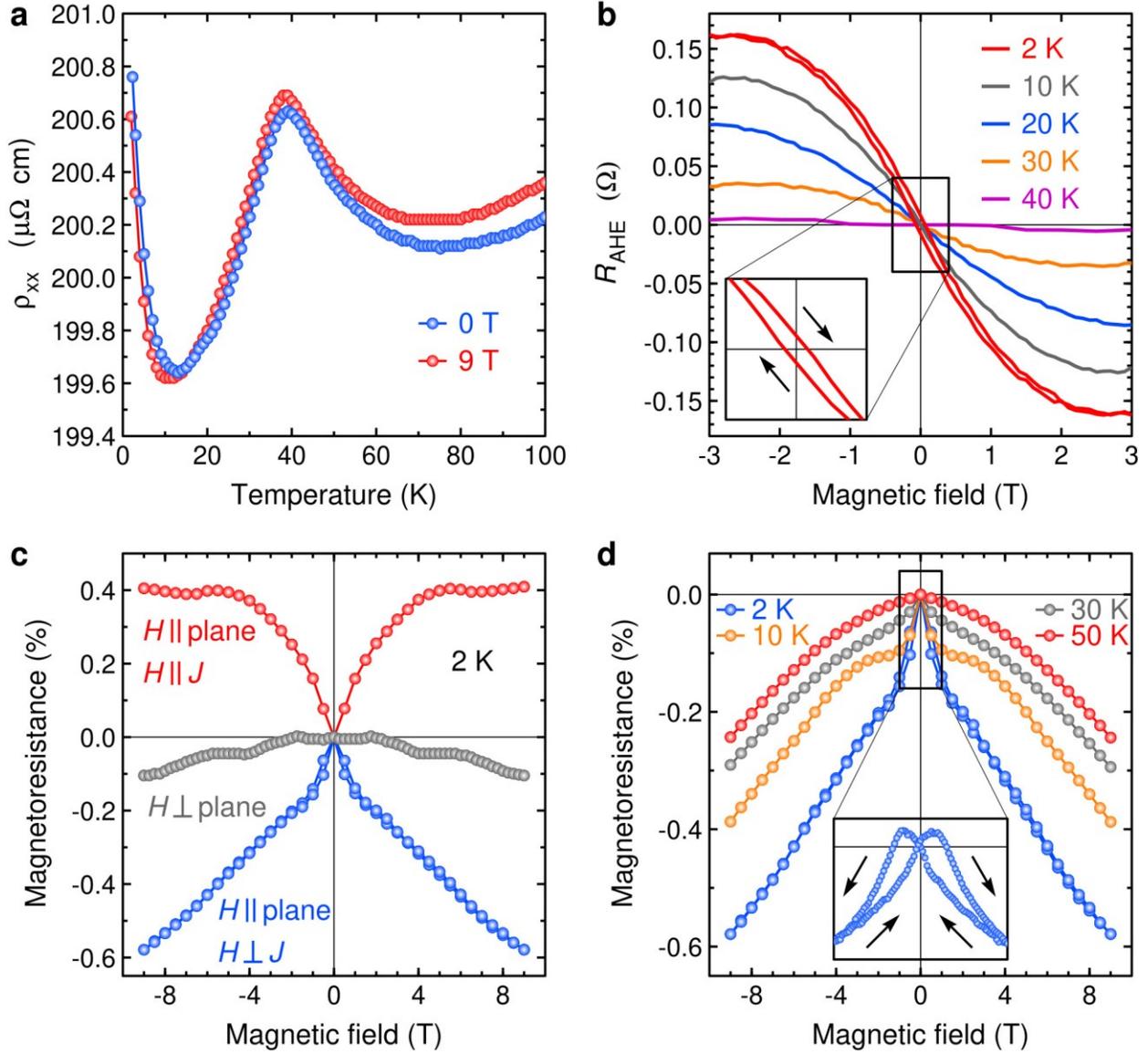

**Figure 6.** Electron transport in the trigonal polymorph of GdAlSi (11 ML). a) Temperature dependence of resistivity in zero field (blue) and an out-of-plane magnetic field 9 T (red); b) magnetic field dependence of the anomalous contribution to the Hall effect at 2 K (red), 10 K (grey), 20 K (blue), 30 K (orange), and 40 K (purple); inset: magnified anomalous Hall effect at 2 K to demonstrate the presence of hysteresis; c) magnetoresistance at 2 K in out-of-plane magnetic fields (grey) and in-plane magnetic fields parallel (red) and perpendicular (blue) to the current; d) magnetoresistance in in-plane magnetic fields perpendicular to the current at 2 K (blue), 10 K (orange), 30 K (grey), and 50 K (red); inset: a detailed study of magnetoresistance at 2 K demonstrating hysteresis at low magnetic fields.